\documentstyle[12pt, epsf]{article}
\begin{document}

\begin{center}
{\large \bf 
Moving Stationary State of 
}
\end{center}
\begin{center}
{\large \bf
Exciton\,-\,Phonon Condensate in Cu$_{2}$O 
} 
\end{center}
\vspace*{0.3cm}
\begin{center}
D. Roubtsov and Y. L\'epine
\end{center}
\begin{center}
{\it 
Groupe de Recherche en Physique et Technologie des Couches Minces, 
}
\end{center}
\begin{center}
{\it 
Departement de Physique, Universit\'e de Montr\'eal, 
}
\end{center}
\begin{center}
C.P. 6128, succ. Centre-ville, Montreal, PQ, H3C\,3J7, Canada 
\end{center}
\begin{center}
e-mail: roubtsod@physcn.umontreal.ca 
\end{center}
\vspace*{0.5cm}
\begin{abstract}

We explore a simple theoretical model to describe 
the properties of Bose condensed para-excitons in  Cu$_{2}$O.
Taking into account the exciton\,-\,phonon interaction and introducing   
a coherent phonon part of the moving condensate, we derive the dynamic 
equations for the exciton\,-\,phonon condensate. 
Within the Bose approximation for excitons, we discuss the conditions
for the moving inhomogeneous condensate to appear in the crystal.   
We calculate the condensate wave function and energy and 
a collective excitation spectrum in the semiclassical approximation.
The stability conditions of the moving condensate are analyzed by use of
Landau arguments, and two critical velocities appear in the theory.
Finally, we apply our model to describe
the recently observed interference between two coherent exciton\,-\,phonon 
packets in Cu$_{2}$O. 
\end{abstract}

\vspace*{0.5cm}


\section{Introduction}

Excitons in semiconductor crystals \cite{review} and nanostructures 
\cite{Butov} are a very interesting and challenging object to search
for the process of Bose Einstein condensation (BEC).  
Nowadays there is a lot of experimental evidence that the optically 
inactive para-excitons in
Cu$_{2}$O can form a highly correlated state, or the excitonic 
Bose Einstein condensate 
\cite{review},\cite{Lin},\cite{Goto}.
A moving condensate of para-excitons in a
3D Cu$_{2}$O crystal turns out to be spatially  inhomogeneous  in the 
direction of
motion, and the registered velocities of coherent exciton packets 
turn out to be always less, but
approximately equal to the longitudinal sound speed of the crystal \cite{Fortin}.

Analyzing recent experimental \cite{Lin},\cite{Fortin},\cite{Benson}
and theoretical \cite{Hanamura1}-\cite{boser}
studies of BEC of excitons in Cu$_2$O, we can conclude that there are
essentially two different stages of this process. The first stage is
the kinetic one, with the characteristic time scale of $10\sim 20$\,ns.
 At this stage, a condensate of long-living para-excitons begins to be 
formed  from a quasi-equilibrium 
degenerate state of excitons ($\mu \ne 0$, $T_{\rm eff} > T_{\rm latt}$) 
when the concentration and the effective temperature 
of excitons in a cloud  meet the conditions of Bose-Einstein Condensation \cite{review}.   
Note that we do not discuss here the behavior of ortho-excitons (with the 
lifetime $\tau_{\rm ortho} \simeq 30$\,ns) and their influence on 
the para-exciton condensation process. For more details about the 
ortho-excitons
in Cu$_2$O, ortho-para-exciton conversion, etc.  see \cite{Lin},\cite{Goto},\cite{Kavoulakis}.  

The most intriguing feature of the kinetic stage is that formation of 
the para-exciton condensate 
and the process of momentum transfer to the para-exciton cloud are
happening simultaneously. It seems that nonequilibrium acoustic phonons
(appearing at the final stage of exciton cloud cooling) play the key role in 
the 
process of momentum transfer. Indeed, the theoretical results obtained in 
the framework of the ``phonon wind'' model \cite{Tichodeev},\cite{discussion} 
and the experimental observations \cite{Lin},\cite{Goto},\cite{Fortin}
are the strong arguments in favor of this idea.
To the authors' knowledge, there are no realistic theoretical 
models of the kinetic stage of para-exciton condensate formation  where  
quantum degeneracy of the initial exciton state and possible coherence of 
nonequilibrium phonons pushing the excitons would be taken into account.
Indeed, the condensate formation and many other processes involving it are essentially
nonliner ones. Therefore, the condensate, or, better, the macroscopically occupied
mode, can be different from $ n({\bf k}=0) \gg 1$, and the language of the states in
${\bf k}$-space and their occupation numbers $ n({\bf k})$ may be not relevant to 
the problem, see \cite{Ivanov}.
 
In this study, we will not explore the stage of condensate 
formation. Instead, we investigate 
the second, quasi-equilibrium stage, in which the condensate has already been 
formed and it moves through a crystal with some constant velocity and 
characteristic shape of the density profile.
In theory, the time scale of this ``transport'' stage,
$\Delta t_{\rm tr}$,
 could be 
determined by the para-exciton lifetime 
($\tau_{\rm para} \simeq 13\,\,\mu$s \cite{review}). 
In practice, it is determined by the characteristic size $\ell$ of a 
high-quality single crystal available for experiments: 
 $$
\Delta t_{\rm tr} \simeq \ell/c_{l}\simeq 0.5-1.5\,\mu{\rm s}   
\ll \tau_{\rm para}, 
$$
where $c_{l}$
is the longitudinal sound velocity. 
        
We assume that at the ``transport'' stage, 
the temperature of the moving packet 
(condensed\,$+$\,noncondensed particles) is equal to the lattice 
 temperature, $$ 
T_{\rm eff}= T_{\rm latt}<T_{c}.$$
Then we can consider first the simplest case of $T=0$ and disregard the influence of all sorts
of {\it nonequilibrium} phonons (which appear at the stages of exciton 
formation, thermalization \cite{Tichodeev}
) on 
the formed moving condensate.

Any theory of the exciton BEC in Cu$_{2}$O has to point out some physical mechanism(s)
by means of which the key experimental facts can be explained. (For example,  
the condensate moves without friction
within a narrow interval of velocities localized near $c_{l}$, 
and   
the shape of the stable macroscopic wave function of excitons
resembles soliton profiles \cite{Benson}.) 
Here we explore a simple model of exciton-phonon condensate.
In this case, the general structure of the Hamiltonian of the moving
exciton packet and the lattice phonons is the following:  
\begin{equation}
\hat{H}=H_{\rm ex}(\hat{\psi}^{\dag},\hat{\psi}) 
        - {\bf v}{\bf P}_{\rm ex}(\hat{\psi}^{\dag},\hat{\psi})
        + H_{\rm ph}(\hat{{\bf u}}, \hat
{\pi}
)
        - {\bf v}{\bf P}_{\rm ph}(\hat{{\bf u}}, \hat
{\pi}
 )
        + H_{\rm int}(\hat{\psi}^{\dag}\hat{\psi},\,\partial_{j}\hat{ u}_{k}). 
 \label{ham1}
\end{equation}
 Here $\hat{\psi}$ is the Bose-field operator describing the excitons,
$\hat{\bf u}$ is the field operator of lattice displacements,  
$\hat
{\pi}
$ 
is the momentum density operator canonically 
conjugate to $\hat{\bf u}$, ${\bf v}$ is the exciton packet velocity and,
finally, 
${\bf P}$ is  the momentum operator. Note 
that the Hamiltonian (\ref{ham1}) is 
written in the reference frame moving with the exciton packet, i.e.
$ {\bf x} \rightarrow {\bf x} - {\bf v}t$ and ${\bf v} = {\rm 
const}$ is the packet velocity.     

\section{3D Model 
of Moving Exciton-Phonon Condensate}

To derive the equations of motion of the field operators (and generalize these equations to the case of  
$T \ne 0 $), it is more convenient, however, to start from the Lagrangian.
In the proposed model, the Lagrangian density has the form
$$
{\cal L}={{i\hbar}\over{2}}( \hat{\psi}^{\dag}\,\partial_{t}\hat{\psi} - 
                    \partial_{t}\hat{\psi}^{\dag}\,\hat{\psi}) 
 +\,v{{i\hbar}\over{2}}(\partial_{x}\hat{
\psi}^{\dag}\,\hat{\psi} - 
\hat{\psi}^{\dag}\partial_{x}\hat{\psi}) -
$$
$$
- E_{g}\hat{\psi}^{\dag}\hat{\psi} - {{\hbar^{2}}\over{2m}}
\nabla
\hat{\psi}^{\dag}\nabla\hat{\psi}
- {{\nu_{0}}\over{2}} \bigl(\psi^{\dag}({\bf x},t)\bigr)^{2}
\bigl(\psi({\bf x},t)\bigr)^{2} - {{\nu_{1}}\over{3}}\bigl(\psi^{\dag}({\bf 
x},t)\bigr)^{3}\bigl(\psi({\bf x},t)\bigr)^{3} +
$$
$$
+ {{\rho}\over{2}}(\partial_{t}\hat{\bf u})^{2} -{{\rho c_{l}^{2}}\over{2}} 
\partial_{j}\hat{u}_{s}\,\partial_{j}\hat{u}_{s} 
- \rho v\partial_{t}\hat{\bf u}\,\partial_{x}\hat{\bf u} +
{{\rho v^{2}}\over{2}}(\partial_{x}\hat{\bf u})^{2}-
$$
\begin{equation}
-\sigma_{0} \hat{\psi}^{\dag}({\bf x},t)\hat{\psi}({\bf x},t)
\nabla\hat{\bf u}({\bf x},t),
\label{lagrangian}
\end{equation}
where  $m$ is the exciton ``bare''mass ($m=m_{e}+m_{h}\simeq 3m_{e}$
for $1$s\,excitons in Cu$_{2}$O),  \,$\nu_{0}$ is the exciton-exciton 
interaction
constant ($\nu_{0}>0$ that corresponds to the repulsive interaction between 
para-excitons in Cu$_{2}$O \cite{Schmitt-Rink}), $\rho$ is the crystal density,  
$\sigma_{0}$ is the exciton-longitudinal phonon coupling constant, 
and ${\bf v}=(v,0,0)$. The energy of a free exciton is $E_{g} + \hbar^{2}{\bf k}^{2}/2m$.   
Although the validity of the condition \cite{Keldysh}
$n\tilde{a}_{\rm B}^{3} \ll 1$ 
($\tilde{a}_{\rm B}$ is the exciton Bohr radius)
makes it possible to disregard
all the multiple-particle interactions with more than two participating 
particles in $\hat{H}_{ex}$, we leave the hard-core repulsion term
originated from the three-particle interaction in $\cal L$, i.e. 
$\nu_{1} \ne 0$ and $0< \nu_{1} \ll \nu_{0}$.  For simplicity's sake, we 
take
all the interaction terms in ${\cal L}$ in the {\it local} form and 
disregard the interaction between the excitons and transverse phonons.     
The  packet velocity
$ v \equiv v_{s}$ is one of the parameters of the theory, and we will not 
take into account  
the excitonic normal component and velocity, $(T=0)$.   

The equations of motion can be easily derived by the standard variational
method from the following condition: 
$$
\delta S = \delta \!\int dt\,d{\bf x}\,
{\cal L}\bigl(\hat{\psi}^{\dag}({\bf x},t),\,\hat{\psi}({\bf x},t), \,\hat{\bf u}({\bf x},t)
\bigr)=0.  
$$
Indeed, after transforming the Bose-fields $\hat{\psi}^{\dag}$ and  
$\hat{\psi}$ by 
$$
\hat{\psi}({\bf x},t) \rightarrow \exp(-iE_{g}t/\hbar)
\exp (imvx/\hbar)\hat{\psi}({\bf x},t),
$$
we can write these equations as follows:
$$
(i\hbar\partial_{t} + mv^{2}/2) \hat{\psi}({\bf x},t)= 
$$
\begin{equation}
=\Bigr(- {{\hbar^{2}}\over{2m}}\Delta + \nu_{0}
\hat{\psi}^{\dag}\hat{\psi}({\bf x},t)+ 
\nu_{1}\hat{\psi}^{\dag \,2}\hat{\psi}^{2}({\bf x},t)\,\Bigl)  
\hat{\psi}({\bf x},t) + \sigma_{0}\nabla\hat{\bf u}({\bf x},t)\,
\hat{\psi}({\bf x},t),
\label{eq1}
\end{equation}
\begin{equation}
\bigl(\partial_{t}^{2}- c_{l}^{2}\Delta - 2v\partial_{t}\partial_{x} 
+ v^{2}\partial_{x}^{2}\bigr)\hat{\bf u}({\bf x},t)=
\rho^{-1}\sigma_{0}\nabla\bigl(\hat{\psi}^{\dag}\hat{\psi}({\bf x},t)\bigr).
\label{eq2}
\end{equation}

We  assume that the condensate of excitons exists. This  means that 
the following representation of the exciton Bose-field holds: 
 $\hat{\psi}=\psi_{0} + \delta\hat{\psi}$.
Here $\psi_{0} \ne 0$ is the classical part of the field operator $\hat{\psi}$
or, in other words, the condensate wave function, and  $\delta\hat{\psi}$
is the fluctuational part of $\hat{\psi}$, which describes noncondensed particles.

One of the important objects in the theory of BEC is the correlation functions
of Bose-fields. The standard way to calculate them in this model (the 
excitonic function $<\!\!\psi({\bf x},t)\psi^{\dag}
({\bf x}',t')\!\!>$, for example,) can be based on the effective 
action or the effective 
Hamiltonian approaches \cite{Popov}.   Indeed, one can, first, integrate over the
phonon variables ${\bf u}$, get the expression for 
$ S_{\rm eff}(\psi, \psi^{\dag})$ and, second, use $S_{\rm eff} $ (or 
$\hat{H}_{\rm eff}$) to derive the equations of
motion for $\psi_{0}$, $\delta\hat{\psi}$, correlation functions, etc..   

In this work  we do not follow  that way; instead, we treat excitons and 
phonons equally \cite{Loutsenko},\cite{miscellaneous}. This means that 
the displacement field $\hat{\bf u}$ can have 
 a nontrivial classical part too, i.e. $\hat{\bf u}={\bf u}_{0} + \delta\hat{\bf u}$ and
${\bf u}_{0} \ne 0$, and the actual moving condensate can be an exciton-phonon one,
i.e. $\psi_{0}({\bf x},t)\cdot {\bf u}_{0}({\bf x},t)$.  
Then the equation of motion for the classical parts of the fields $\hat{\psi}$ and $\hat{\bf u}$
can be derived by use of the variational method from 
${\cal L} ={\cal L}(\psi, \psi^{*}, {\bf u})$, where all the fields can be considered as the classical ones.
Eventually we have 
$$
(\,i\hbar\partial_{t} + mv^{2}/2\,){\psi}_{0}({\bf x},t)= 
$$
\begin{equation}
=\Bigr(- {{\hbar^{2}}\over{2m}}\Delta + \nu_{0}
\vert{\psi_{0}}\vert^{2}({\bf x},t)+ 
\nu_{1}\vert{\psi}_{0}\vert^{4}({\bf x},t)\,\Bigl)  
{\psi}_{0}({\bf x},t) + \sigma_{0}\nabla{\bf u}_{0}({\bf x},t)\,
{\psi}_{0}({\bf x},t)\,,
\label{eq11}
\end{equation}
\begin{equation}
\bigl(\partial_{t}^{2}- c_{l}^{2}\Delta - 2v\partial_{t}\partial_{x} 
+ v^{2}\partial_{x}^{2}\bigr){\bf u}_{0}({\bf x},t)=
\rho^{-1}\sigma_{0}\nabla\bigl(\vert\psi_{0}\vert^{2}({\bf x},t)\bigr).
\label{eq22}
\end{equation}
Notice that deriving these equations we disregarded the interaction between
the classical (condensate) and the fluctuational 
(noncondensate) parts of the fields. That is certainly a good approximation
for $T=0$ and $T \ll T_{c}$ cases \cite{Griffin}.

In this article a steady-state of the condensate is the object of the main 
interest. In the co-moving frame of reference, 
the condensate steady-state is just the  stationary solution of Eqs. 
(\ref{eq11}),\,(\ref{eq22}) and it can be taken in the form
$$
\psi_{0}({\bf x},t)=\exp(-i\mu t)\exp(i\varphi)\phi_{\rm o}({\bf x}), 
\,\,\,\,\,
u_{0}({\bf x},t)= {\bf q}_{\rm o}({\bf x}),
$$ 
where  $\phi_{\rm o}$ and  ${\bf q}_{\rm o}$
are the real number functions, and $\varphi ={\rm const}$ is the (macroscopic) phase
of the condensate wave function. (This phase can be taken zeroth if only a single condensate 
is the subject of interest.)
  Then,  the following equations have to be solved: 
\begin{equation}
\tilde{\mu}{\phi}_{\rm o}({\bf x})= 
\Bigr(- {{\hbar^{2}}\over{2m}}\Delta + \nu_{0}
\phi_{\rm o}^{2}({\bf x})+ 
\nu_{1}\phi_{\rm o}^{4}({\bf x})\,\Bigl)  
\phi_{\rm o}({\bf x}) + \sigma_{0}\nabla{\bf q}_{\rm o}({\bf x})\,
{\phi}_{\rm o}({\bf x})\,,
\label{eq111}
\end{equation}
\begin{equation}
- \bigl(\,(c_{l}^{2}- v^{2})\partial_{x}^{2}
+ c_{l}^{2}\partial_{y}^{2} + c_{l}^{2}\partial_{z}^{2}\,\bigr)
{\bf q}_{\rm o}({\bf x})=
\rho^{-1}\sigma_{0}\nabla\phi_{\rm o}^{2}({\bf x}).
\label{eq222}
\end{equation}
Indeed, the last equation can be solved relative to $\nabla {\bf 
q}_{\rm o}$. 
If $v<c_{l}$, the corresponding solution can be represented as follows:
\begin{equation}
\nabla{\bf q}_{\rm o}({\bf x})=\bigl(-\frac{\lambda^{2}}{3}-\frac{2}{3}\bigr)
{{\sigma_{0}}\over{\rho 
c_{l}^{2}}}\,\phi_{\rm o}^{2}({\bf x})\,+\, 
{\lambda^{3}\over{4\pi}}(1-\lambda^{-2})\int\! {\cal F}({\bf x}-{\bf 
x}')\frac{\sigma_{0}}{\rho c_{l}^{2}}\,\phi^{2}_{\rm o}({\bf x}')\,d{\bf 
x}',  
\label{phonon1}
\end{equation}
where $\lambda^{2}=c_{l}^{2}/(c_{l}^{2} - v^{2})$, and ${\cal F}$ can be 
expressed in terms of the Green function of Eq. 
(\ref{eq222}). 
Substituting $\nabla{\bf q}_{\rm o}$ in Eq. 
(\ref{eq111}), we rewrite the latter in the following form:
\begin{equation}
\tilde{\mu}\phi_{\rm o}({\bf x})= \Bigr(-{{\hbar^{2}}\over {2m}}\Delta +
\nu_{0}\phi_{\rm o}^{2}({\bf x})
+\int\!U_{\rm eff}({\bf x}-{\bf x}')
\phi_{\rm o}^{2}({\bf x}')\,d{\bf x}' + \nu_{1}\phi_{\rm o}^{4}({\bf 
x})\,\Bigl)\phi_{\rm o}({\bf x}), 
\label{3Deq}
\end{equation}
where the effective exciton-exciton interaction $U_{\rm eff}$ is induced by 
the lattice ($q_{\rm o}\ne 0$). It can be represented as follows:
$$
U_{\rm eff}({\bf x})= \Bigl(-\frac{\lambda^{2}}{3}-\frac{2}{3}\Bigr){ 
{\sigma_{0}^{2}} \over {\rho c_{l}^{2}} }\delta({\bf x}) +
$$
\begin{equation}
+  { {\sigma_{0}^{2}}\over 
{4\pi\rho (c_{l}^{2} - v^{2} )} }\,{ {v^{2}} \over
 {c_{l}(c_{l}^{2}-v^{2})^{1/2}} } \Bigr(\,
\frac{3\lambda^{2}x^{2}} {(\lambda^{2}x^{2}+ y^{2}+ z^{2})^{5/2}} -
\frac{1}{(\lambda^{2}x^{2} + y^{2} + z^{2})^{3/2} } \,\Bigl).  
\label{potential}
\end{equation} 
The first (isotropic) term in (\ref{potential}) causes the renormalization
of the exciton-exciton interaction constant $\nu_{0}> 0$: 
\begin{equation}
\nu_{0}\rightarrow \nu_{\rm eff}=\nu(v;\,c_{l}, \sigma_{0}). 
\label{renorm}
\end{equation}
Note that $\nu_{\rm eff}$ can be positive or negative depending on the
value of $v$.  
The second term in the effective potential (\ref{potential}) is strongly 
anisotropic. Moreover, on the cylinder $y^{2}+z^{2}=\varepsilon^{2}$, 
its value is negative in the vicinity of $x=0$
and positive  at the large scales of $x$ \,(${\rm O}x \| {\bf v}$).
 
The possibility of the existence of specific and to some extent unexpected
solutions of Eq. (\ref{3Deq}) follows  from the fact that 
the effective two particle interaction between excitons can be attractive  
at small distances between the particles and repulsive at large distances. 
 For example, the  wave function $\phi_{\rm o}({\bf x})$  
may become strongly localized in 3D space because of this attraction.  

If $v > c_{l}$,
the formulas for $\nabla{\bf q}_{\rm o}({\bf x})$ and \,$U_{\rm eff}({\bf x})$  are very
similar to those obtained in the case of $v<c_{l}$.  For instance,
the solution of Eq. (\ref{eq222}) can be written as follows:
\begin{equation}
\nabla{\bf q}_{\rm o}({\bf x})=\bigl(\frac{\tilde{\lambda}^{2}}{3}-\frac{2}{3}\bigr)
{ {\sigma_{0}} \over {\rho 
c_{l}^{2}} }\,\phi_{\rm o}^{2}({\bf x})\,+\, 
\frac{\tilde{\lambda} }{2\pi}(1+\tilde{\lambda}^{2})\int\! {\cal F}'({\bf x}-{\bf 
x}')\frac{\sigma_{0}}{\rho c_{l}^{2}}\,\phi^{2}_{\rm o}({\bf x}')\,d{\bf 
x}'. 
 \label{phonon101}
\end{equation}
(Here $\tilde{\lambda}^{2} = c^{2}_{l}/(v^{2}-c^{2}_{l})$).
Again, the renormalization (\ref{renorm}) takes place and 
a nonisotropic part of $U_{\rm eff}'({\bf x})$ appears in  Eq. (\ref{3Deq}).
However, the effective two-particle interaction between excitons remains 
repulsive, i.e. $\nu_{0}\delta({\bf x}) + U_{\rm eff}'({\bf x})>0$.

\section{Effective 1D Model for the Condensate Wave Function}

Solving Eqs. (\ref{eq111}),(\ref{eq222})
in 3D space  seems to be a difficult problem (see  
Eqs. (\ref{3Deq}),(\ref{potential})\,).
However, these equations can be essentially simplified  if we 
assume that the condensate is inhomogeneous along the 
$x$-axis only, that is  
$\phi_{\rm o}({\bf x})=\phi_{\rm o}(x)$ 
and ${\bf q}_{\rm o}({\bf x})=(q_{\rm o}(x), \,0,0)$.
Such an effective reduction of dimensionality transforms the difficult  
integro-differential equation (\ref{3Deq}) into a rather simple  
differential one, and obtained in this way the    
 effective 1D model for the condensate wave function  
$\phi_{\rm o}\circ q_{\rm o}$ \,conserves all the important properties       
of the ``parent'' 3D model. Indeed, if $v<c_{l}$, the following equations 
stand for the condensate:
\begin{equation} 
\tilde{\mu}\phi_{\rm o}(x)=\bigl(\, -(\hbar^{2}/2m) 
\partial_{x}^{2} +\nu_{\rm eff}\phi_{\rm o}^{2}(x) +
\nu_{1}\phi_{\rm o}^{4}(x)\,\bigr)\phi_{\rm o}(x),
\label{1Deq} 
\end{equation} 
\begin{equation}
\partial_{x} q_{\rm o}(x)={\rm const} -\bigl( \sigma_{0}/\rho
(c_{l}^{2}-v^{2})\bigr)\,\phi_{\rm o}^{2}(x),
\label{11Deq}
\end{equation}
where $\nu_{\rm eff}=\nu_{0}-\sigma_{0}^{2}/\rho(c_{l}^{2}-v^{2})$.
If $v>c_{l}$, Eq. (\ref{1Deq}) describes the excitonic part of the condensate, 
but  with the enhanced effective repulsion 
$\nu_{\rm eff}'=\nu_{0}+\sigma_{0}^{2}/\rho (v^{2}-c_{l}^{2})$. 

The effective two-particle interaction constant $\nu_{\rm eff}$ is  negative
if the velocity of the condensate lies inside the interval $v_{\rm o}<v<c_{l}$, 
 where 
\begin{equation}
v_{\rm o}=\sqrt{c_{l}^{2}-(\sigma_{0}^{2}/\rho\nu_{0}) }
\label{velo1}
\end{equation}
can be called  the first `critical' velocity in the model.
(Note that outside this interval $\nu_{\rm eff}>0$ \cite{Loutsenko}.)

The estimate value of the threshold velocity $v_{\rm o}$ can be  obtained from the following formula:
$$
v_{\rm o} \simeq c_{l}\,\sqrt{1\,-\,(C_{c}-C_{v})^{2}/(8\pi\,{\rm Ry}^{*}\,{\rm Ry}\gamma^{3}) },
$$
where $C_{c}-C_{v}$ is the relative volume deformation potential of a semiconductor, 
($\sigma_{0}\simeq C_{c}-C_{v}$), ${\rm Ry}^{*}$ and Ry are
the exciton and atom Rydberg  energies, $\gamma= \tilde{a}_{\rm{B}}/a_{l}$ 
and  $a_{l}$ is the lattice constant. (The repulsive exciton-exciton interaction is taken 
in the form $\nu_{0}\simeq 4\pi \hbar^{2}a_{\rm{ex}}/m$). In the case of Cu$_{2}$O oxide,
we have $v_{\rm o}\simeq (0.5\sim 0.7)c_{l}$.

 In this study we will consider the case of $v_{\rm o}<v<c_{l}$ in 
detail.
If the sign of the effective two-particle interaction can vary, an  
extra (repulsive) term should be included into the Hamiltonian of a 
many-particle system to insure 
the finiteness of a steady-state wave function or the absence of wave 
function collapse
in dynamic processes. 
In our case, it is the term $\nu_{1}(\Psi^{\dag})^{3}\Psi^{3}$, and 
$\nu_{1}>0$ is supposed to be the smallest energy parameter in the theory.
In the framework of the considered 1D model, however, 
a finite steady-state solution of Eqs. (\ref{1Deq}),(\ref{11Deq}) can be 
obtained without  
accounting 
for  
the ``hard core'' repulsion term $\nu_{1}\phi_{\rm o}^{4}$.   
 Indeed, we can write out the corresponding solution as follows:
\begin{equation}
\phi_{\rm o}(x)=\Phi_{\rm o}\cosh^{-1}(\beta\Phi_{\rm o}x), \,\,\,\,
\partial_{x}q_{\rm o}(x)=
-\bigl(\sigma_{0}
/\rho(c_{l}^{2}-v^{2})\bigr)\Phi_{\rm o}^{2}\cosh^{-2}(\beta\Phi_{\rm o}x),
 \label{Soliton}
\end{equation}
\begin{equation}
\tilde{\mu}=\nu_{0}\Phi_{\rm o}^{2}/2\,-\, 
\bigl(\sigma_{0}^{2}/\rho(c^{2}_{l}-v^{2})\bigr)\Phi_{\rm o}^{2}/2 < 0,
 \end{equation}
 \begin{equation} 
\beta=\beta(v)=\sqrt{\frac{m\nu_{0}}{\hbar^{2}} 
\frac{v^{2}-v_{\rm o}^{2}}{c_{l}^{2}-v^{2}} }.
\end{equation}

The amplitudes of the exciton and phonon parts of the condensate, 
the characteristic width of the 
condensate, $L_{0}=(\beta (v)\Phi_{\rm o})^{-1}$,  \,and the value of the 
effective chemical 
potential $\tilde{\mu}$ depend on the normalization of the exciton wave 
function $\phi_{\rm o}(x)$. We normalize it in 3D space assuming that the 
characteristic width of the packet in the $(y,z)$ plane is finite and
the cross-section area of the packet can be made equal to  
the cross-section area S of a laser beam. Then we can write this condition as follows:   
\begin{equation}
\int\!\!\vert\psi_{0}\vert^{2}(x,t)\,d{\bf x}
=S\int\!\!\phi_{\rm o}^{2}(x)\,dx = N_{\rm o},
\label{norma}   
\end{equation} 
where $N_{\rm o}$ is the number of condensed excitons.
Immediately, we get the following results: 
\begin{equation}  
\Phi_{\rm o}=\frac{N_{\rm o} }{2S}\beta(v),\,\,\,\,
L_{0}=\Bigl(\frac{N_{\rm o}}{2S}\beta^{2}(v)\Bigl)^{-1}, 
\label{amplitude}
\end{equation}
\begin{equation}
\tilde{\mu}=-\frac{\nu_{0} }{2}\frac{v^{2}-v_{\rm o}^{2}}{ c_{l}^{2}-v^{2}} 
\Bigl(\frac{N_{\rm o}}{2S}
\beta (v)\Bigr)^{2}=- \frac{\hbar^{2}}{2m}\,L_{0}^{-2}.
\label{mu}
\end{equation}
The important ``nonlinear'' property of the obtained solution (\ref{Soliton}) is the 
dependence
of the amplitudes of exciton and phonon parts of the condensate 
on the velocity $v$ and 
the number of excitons in the condensate, $N_{\rm o}$  . For example, 
$$
\Phi_{\rm o}^{2}=
\Phi_{\rm o}^{2}(N_{\rm o},v)\sim N_{\rm o}^{2}\,(v^{2} - v_{\rm o}^{2})
/(c_{l}^{2} - v^{2})  
$$
stands for the exciton amplitude.
Notice that the characteristic width of the condensate and its velocity are not independent
parameters, see (\ref{amplitude}).  
For estimates, the formula for $L_{0}$ can be rewritten as follows:
$$
L_{0}^{-1}\simeq 2\tilde{a}_{\rm{B}}^{-1}(n\tilde{a}_{\rm {B}}^{3})^{1/2} 
\left(\frac{v^{2}-v_{\rm o}^{2}}{c_{l}^{2}-v^{2}}\right)^{1/2},
$$
where $n$ is the average density of excitons in the soliton state.
Although $\tilde{a}_{\rm{B}}^{-1}$  in this formula is multiplied by a small factor, this factor
is  not small enough to show quantitative agreement with the experimentally observed value
of $L_{0}$, $2L_{\rm exp} \simeq  10^{-1}\sim 10^{-2}\,{\rm cm}$.  
It seems to be reasonable that a theory
with nonzero temperature (\,$T>\vert\tilde{\mu}\vert $\,)
will provide a more realistic value of the effective size of the packet.

Returning to the laboratory reference frame, we can write the condensate wave function 
in the form:  
$$ 
\psi_{0}(x,t)\cdot u_{0}(x,t)\delta_{1j}=
\exp\left(-i\left(E_{g}+\frac{mv^{2}}{2} -\vert\tilde{\mu}\vert \right )t \,\right)\exp(imvx)\times
$$
\begin{equation}
\times \Phi_{\rm o}\cosh^{-1}\bigl( L_{0}^{-1}(x-vt)\,\bigr )\cdot
\Bigl( Q_{\rm o}\,-\,Q_{\rm o}\tanh \bigl(L_{0}^{-1}(x-vt)\bigr)\, \Bigr ),
\label{movingcond}
\end{equation}
where we count the exciton energy from the bottom of the crystal valence band;  
 $2Q_{\rm o}(N_{\rm o},v)$ is the amplitude of the phonon part of 
condensate and $Q_{\rm o} \propto \Phi_{\rm o}$.   
To calculate the energy of the moving condensate within the Lagrangian approach, 
(see Eq. (\ref{lagrangian})\,), we have to integrate the zeroth component
of the energy-momentum tensor ${\cal{T}}_{0}^{0}$ over the spatial coordinates, 
$$
{\cal{T}}_{0}^{0}({\bf x},t)= E_{g}\psi^{\dag}\psi + 
\frac{\hbar^{2}}{2m}\nabla\psi^{\dag}\,\nabla\psi + (\nu_{0}/2)(\psi^{\dag})^{2}\psi^{2}
+ \frac{\rho}{2}(\partial_{t}{\bf u})^{2} +\frac{\rho c_{l}^{2}}{2}\,\partial_{j} u_{k}\partial_{j} u_{k} +
\sigma_{0}\psi^{\dag}\psi\,\nabla {\bf u}. 
$$
We do not take into account the small correction to this energy due to the quantum 
depletion of the condensate as well as the term $(\nu_{1}/3)\,\phi_{\rm o}^{6}$ in ${\cal{T}}_{0}^{0}$.
Then the result reads: 
$$
E _{\rm{o}}=\int\!d{\bf r}\,{\cal{T}}_{0}^{0} = E_{\rm ex} + E_{\rm int} + E_{\rm ph}=
$$
$$
=N_{\rm o}\left(E_{g} + \frac{mv^{2}}{2}\,\right) - 
N_{\rm o}\left(\vert\tilde{\mu}\vert + (\nu_{0}/3)\Phi_{\rm o}^{2}
\,\right)
+
\frac{c_{l}^{2}+v^{2}}{3(c_{l}^{2}-v^{2})}\,N_{\rm o}\frac{\sigma_{0}^{2}}{\rho(c_{l}^{2}-v^{2})}\Phi_{\rm o}^{2}.
$$
The value $\vert \tilde{\mu}\vert $ 
is a rather small parameter,
$$ 
\vert \tilde{\mu}\vert 
\simeq 6{\rm Ry}^{*} (na_{\rm ex}^{3})\,(v^{2}-v_{0}^{2})/(c^{2}_{l}-v^{2}),
$$
and  the energy of the phonon part of the condensate is estimated as 
$E_{\rm 
ph} \le (5\sim 6) N_{\rm o}\vert \tilde {\mu}\vert$.  
However, the back surface of the crystal will experience some pressure
when the condensate reaches this surface and the excitons are destroyed 
near it. The estimation of the maximum value of the pressure in a pulse 
is as follows: 
$$
p_{m} \simeq 10 \sigma_{0} \Phi_{\rm o}^{2} \simeq 10^{-2}\sim 10^{-3}\,{\rm J}/{\rm cm}^{-3} 
$$
and the main contribution to this value comes from the phonon 
part.
Indeed, one can see that the exciton-phonon condensate
carries a nonzeroth momentum $P_{{\rm o}\,x }= P_{{\rm ex},\,x}+ P_{{\rm ph},\,x}$:
$$
P_{{\rm o}\,x}=
\int d{\bf x} (\hbar / 2i)(\phi_{0}^{*}(x,t)\,\partial_{x}\phi_{0}(x,t)-
\partial_{x}\phi_{0}^{*}(x,t)\,\phi_{0}(x,t)\,) - \rho\partial_{t}u_{0}(x,t)\,\partial_{x}u_{0}(x,t)=
$$
$$
=\int d{\bf x}\,mv\,\Phi_{\rm o}^{2}(x) +\rho v \left ( \frac{\sigma_{0} }{(c_{l}^{2}-v^{2})\rho}
\Phi_{\rm o}^{2}(x)\,\right)^{2}.
$$

\section{ Low-Lying Excitations of Exciton-Phonon Condensate}

Although the condensate wave function $\phi_{\rm o}(x)\cdot q_{\rm o}(x)$ was 
obtained 
in the framework of the effective 1D model, (but normalized in 3D space), we will use 
this solution as a classical part in the 3D field operator decomposition:    
\begin{equation}
 \hat{\psi}({\bf x},t)=\exp(-i\mu t)(\phi_{\rm o}(x)+ \delta\hat{\psi}
({\bf x},t)\,),
\label{hatpsi}
\end{equation}
\begin{equation}
\hat{u}_{j}({\bf x},t) =q_{\rm o}(x)\delta_{1j} + \delta\hat{u}_{j}({\bf 
x},t), 
\label{hatu}
\end{equation}
where $\mu=\tilde{\mu} - mv^{2}/2 $. \,Substituting 
the field operators of the form (\ref{hatpsi}),(\ref{hatu}) 
into the Lagrangian density (\ref{lagrangian}), we can write the 
later in the following form:
\begin{equation}
{\cal L}={\cal L}_{\rm o}({\rm e}^{-i\mu t}\phi_{\rm o}(x),\,q_{\rm 
o}(x)\delta_{1j}\,)\, +\, 
{\cal L}_{2}(\delta\hat{\psi}^{\dag}({\bf x},t), 
\delta\hat{\psi}({\bf x},t), \,\delta \hat{\bf u}({\bf x},t)\,)\,+... ,  
\label{lagrangian1}
\end{equation}  
where ${\cal L}_{\rm o}$ stands for the classical part of ${\cal L}$, and 
${\cal L}_{2}$ is the bilinear form in the $\delta$-operators.
As the classical parts of the field operators satisfy the equality $\delta 
S_{\rm o}[\psi_{0}^{*},\psi_{0},\,u_{0}]=0$, 
the linear form in the ``$\delta$-operators''  vanishes in 
(\ref{lagrangian1}).

In the simplest (Bogoliubov) approximation \cite{Fetter},\cite{Pit},
\,${\cal L}\approx {\cal 
L}_{\rm o}+ {\cal L}_{2}$ and, hence, the bilinear form ${\cal L}_{2}$ 
defines the equations of motion for the fluctuating parts of the field 
operators. (To derive them we use the  variational method:  
$
\delta S_{2}=\delta\int\!{\cal L}_{2}(\delta\psi^{\dag}, \delta\psi,
\delta {\bf u})\, d{\bf x}dt = 0
$.)   
As a result, these equations are 
linear and can be written as follows:     
$$
i\hbar\partial_{t}\,\delta\hat{\psi}({\bf x},t)=
\left(-\frac{\hbar^{2}}{2m}\Delta + \vert\tilde{\mu}\vert +
\left\{ 2\nu_{0}- \frac{\sigma_{0}^{2}}{\rho (c_{l}^{2}-v^{2}) }\right \}
\phi_{\rm o}^{2}(x) +3\nu_{1}\phi_{\rm o}^{4}(x)\,\right)
\delta\hat{\psi}({\bf x},t)+ 
$$
\begin{equation}
+ (\nu_{0}\phi_{\rm o}^{2}(x) + 2\nu_{1}\phi_{\rm o}^{4}(x))\delta\hat{\psi}^{\dag}({\bf x},t)
+ \sigma_{0}\phi_{\rm o}(x)\nabla \delta\hat{\bf u}({\bf x},t),
\label{deltapsi}
\end{equation}
\begin{equation}
\bigl(\partial_{t}^{2}-c_{l}^{2}\Delta -2v\partial_{t}\partial_{x} +v^{2}\partial_{x}^{2}\bigr)
\delta\hat{\bf u}({\bf x},t)=\rho^{-1}\sigma_{0}\nabla\left( \phi_{\rm o}(x)\left(\delta\hat{\psi}({\bf x},t) 
+\delta\hat{\psi}^{\dag}({\bf x},t)\right )\,\right)
\label{deltau}
\end{equation}

The same approximation can be performed within the Hamiltonian approach.
Indeed,  decomposition of the field operators near 
their nontrivial classical parts
leads to the decomposition of the Hamiltonian (\ref{ham1}) itself, and -- 
as it was done with the Lagrangian -- 
only the classical part of $\hat{H}$, $H_{\rm o}$, and  
the bilinear form in the fluctuating fields, $\hat{H}_{2}$, are left for examination: 
\begin{equation}
\hat{H} \approx  H_{\rm o}( \psi_{0}^{*}, \psi_{0}, \,\pi_{0}, u_{0})+
H_{2}(\delta\hat{\psi}^{\dag},\,\delta\hat{\psi}, \,\delta\hat{\pi}, 
\delta\hat{u}).
\label{ham2}
\end{equation} 
In this approximation, the Hamiltonian (\ref{ham2}) can be diagonalized
and rewritten in the form:  
\begin{equation}
\hat{H}=H_{\rm o}({\rm e}^{-i\mu t}\phi_{\rm o}(x),\,q_{\rm o}(x)\,)
+\delta E_{\rm o} + 
\sum_{s}\hbar\omega_{s}\, \hat{\alpha}^{\dag}_{s}\hat{\alpha}_{s}.
\label{ham3}
\end{equation}
Here, $\delta E_{\rm o}$ is the quantum correction to the energy of
the condensate and the index ${s}$ labels the elementary excitations of the system. 
The operators $\hat{\alpha}^{\dag}_{s}$, 
$\hat{\alpha}_{s}$ are the Bose ones, and they can be represented by
linear combinations of the exciton and displacement field 
operators:
\begin{equation}
\hat{\alpha}_{s}=\int \!\!d{\bf x} \left(\,U_{s}({\bf x})\,\delta\hat{\psi}({\bf x})+
                 V_{s}^{*}({\bf x})\,\delta\hat{\psi}^{\dag}({\bf x}) 
+X_{s,j}({\bf x})\,\delta\hat{u}_{j}({\bf x})
+Y_{s,j}({\bf x})\,\delta\hat{\pi}_{j}({\bf x})\,\right)
\end{equation}
\begin{equation}
\hat{\alpha}_{s}^{\dag}=\int\!\! d{\bf x} \left(\,U_{s}^{*}({\bf x})\,\delta\hat{\psi}^{\dag}({\bf x})+
                 V_{s}({\bf x})\,\delta\hat{\psi}({\bf x}) 
+X_{s,j}^{*}({\bf x})\,\delta\hat{u}_{j}({\bf x})
+Y_{s,j}^{*}({\bf x})\,\delta\hat{\pi}_{j}({\bf x})\,\right)
\end{equation}
Note that by analogy with the exciton-polariton modes in semiconductors\,\cite{Hopfield}
the excitations 
of the condensate (\ref{movingcond}) can be considered as a mixture of 
the exciton- and phonon-type modes, but in this model the phonons come from fluctuations
of the $u_{0}(x,t)$-part of the condensate.  

Since the $\alpha$-operators (see (\ref{ham3})\,) evolve in time as simply as    
$$
\hat{\alpha}_{s}(t)={\rm e}^{-i\omega_{s}t}\hat{\alpha}_{s}\,\,\,\,
\hat{\alpha}_{s}^{\dag}(t) = 
{\rm e}^{i\omega_{s}t}\hat{\alpha}_{s}^{\dag},
$$
these operators (and the frequencies $\{ \omega_{s} \}$) are 
the eigenvectors (and, correspondingly, the eigenvalues) of the equations of 
motion (\ref{deltapsi}),(\ref{deltau}) obtained within the 
Lagrangian method.  
Then, the time dependent ``$\delta$-operators'' in 
 (\ref{deltapsi}),(\ref{deltau}) can be 
represented by the following linear combinations of the $\alpha$-operators: 
\begin{equation}
\delta\hat{\psi}({\bf x},t)=\sum_{s}{\rm u}_{s}({\bf x})
                             \,\hat{\alpha}_{s}{\rm e}^{-i\omega_{s}t}
\,+\,{\rm v}_{s}^{*}({\bf x})\,\hat{\alpha}_{s}^{\dag}{\rm e}^{i\omega_{s}t},
\label{uvtransform1}
\end{equation}
\begin{equation}
\delta\hat{\psi}^{\dag}({\bf x},t)=\sum_{s}{\rm u}^{*}_{s}({\bf x})
                             \,\hat{\alpha}^{\dag}_{s}{\rm e}^{i\omega_{s}t}
\,+\,{\rm v}_{s}({\bf x})\,\hat{\alpha}_{s}{\rm e}^{-i\omega_{s}t},
\label{uvtransform11}
\end{equation}
\begin{equation}
\delta\hat{u}_{j}({\bf x},t)=\sum_{s}C_{s,j}({\bf x})\,\hat{\alpha}_{s}
{\rm e}^{-i\omega_{s}t}+ C_{s,j}^{*}({\bf x})\,\hat{\alpha}^{\dag}_{s}{\rm e}^{i\omega_{s}t}.
\label{uvtransform2}
\end{equation}
Substituting this ansatz (which is a generalization of the u-v Bogoliubov
transformation) into Eqs. (\ref{deltapsi}),(\ref{deltau}), 
we obtain the following coupled eigenvalue equations \cite{Loutsenko}:  
\begin{equation}
(\hat{L}(\Delta) - \hbar\omega_{s})\,{\rm u}_{s}({\bf x}) + \left(\nu_{0}\phi_{\rm o}^{2}( x) + 
2\nu_{1}\phi_{\rm o}^{4}( x)\right)
{\rm v}_{s}({\bf x}) + \sigma_{0}\phi_{\rm o}( x)\nabla{\bf C}_{s}( x)=0
\label{eigfunction1}
\end{equation}
\begin{equation}
\left(\nu_{0}\phi_{\rm o}^{2}( x) + 
2\nu_{1}\phi_{\rm o}^{4}( x)\right){\rm u}_{s}({\bf x})+
(\hat{L}(\Delta) +\hbar\omega_{s})\,{\rm v}_{s}({\bf x}) +
 \sigma_{0}\phi_{\rm o}( x)\nabla{\bf C}_{s}({\bf x})=0
\label{eigfunction2}
\end{equation}
\begin{equation}
-\rho^{-1}\sigma_{0}\nabla\Bigl(\phi_{\rm o}( x)\,{\rm u}_{s}({\bf x})\Bigr)
-\rho^{-1}\sigma_{0}\nabla\Bigl(\phi_{\rm o}( x)\,{\rm v}_{s}({\bf x})\Bigr)+
\left[(-i\omega_{s}-v\partial_{x})^{2} - c_{l}^{2}\Delta \right]{\bf C}_{s}({\bf x})=0,
\label{eigfunction3} 
\end{equation}
where  
$
\hat{L}(\Delta)= (-\hbar^{2}/2m)\Delta +\vert\tilde{\mu }\vert + 
\left[\,2\nu_{0} - \sigma_{0}^{2}/\rho(c_{l}^{2}-v^{2})\,\right]\phi_{\rm o}^{2}( x)+
3\nu_{1}\phi_{\rm o}^{4}(x).
$

To simplify investigation of the characteristic properties of the different possible 
solutions of  Eqs. 
(\ref{eigfunction1})-(\ref{eigfunction3}),
we  subdivide the excitations (\ref{uvtransform1})-(\ref{uvtransform2}) into two major parts, the 
{\it inside}-excitations  and the {\it outside}-ones.
The {\it inside}-excitations
are  
localized merely inside the packet area, i.e. $\vert {\bf x} \vert < L_{0}$ and 
$\phi_{\rm o}^{2}(x)\approx {\rm const}$,  
 whereas the {\it outside}-excitations
propagate merely in the outside area, i.e. $\vert {\bf x}\vert > (1\sim 2)L_{0}$
and $\phi_{\rm o}^{2}(x) \simeq \Phi_{\rm o}^{2} \exp(-2\vert x \vert /L_{0})
\rightarrow 0$. 
\begin{figure}
\begin{center}
\leavevmode
\epsfxsize = 235pt
\epsfysize = 235pt
\epsfbox{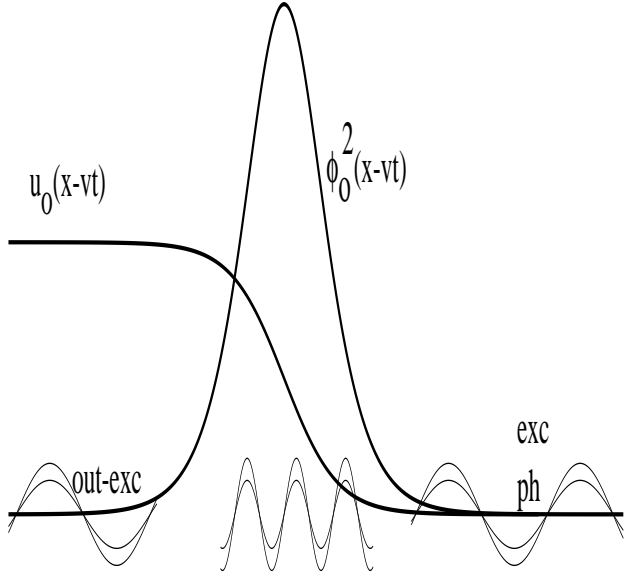}
\end{center}
\caption{
Moving exciton-phonon condensate, $\phi_{\rm o}(x-vt)\cdot u_{\rm o}(x-vt)\delta_{1j}$, and inside- and
outside-excitations of the condensate. (Longitudinal exciton-phonon excitations,
${\bf k}\,||\, Ox$, 
are schematically depicted.)      
}
\end{figure}

\subsection{Outside-Excitations}

For the outside collective excitations, the asymptotics of the low-lying  
energy spectrum can be found easily.  Indeed, if we assume that       
$\phi_{\rm o}^{2}(x) \approx 0$ in the outside packet area, the equations 
(\ref{deltapsi})
and (\ref{deltau}) begin to be uncoupled. 
Then, Eq. (\ref{deltapsi}) describes the excitonic branch  
of the outside-excitations with the following dispertion low in the co-moving frame:
\begin{equation}
\hbar\omega_{{\rm ex}}({\bf k})\approx \vert\tilde{\mu}\vert + (\hbar^{2}/2m)k^{2},\,\,\,\,\, 
({\rm u}_{\bf k}({\bf x})\sim {\rm e}^{i{\bf k}{\bf x}},\,\,\,
{\rm v}_{\bf k}({\bf x})\approx 0),
\label{gap1}
\end{equation}
and Eq. (\ref{deltau}) describes the phonon branch and yields the spectrum 
$\omega_{ph}({\bf k})=c_{l}\vert {\bf k} \vert $ in 
the laboratory frame of reference. 
Then, the exciton field operator, which describes the exciton condensate 
with the one outside excitation,  
has the following form:
$$
\psi({\bf x}, t)\simeq \exp\bigl(-i(E_{g}+mv^{2}/2-\vert\tilde{\mu}\vert)t\,\bigr)\exp(imvx)\,\phi_{\rm o}(x-vt)+ 
$$
$$
+\exp(-i(E_{g}+mv^{2}/2-\vert\tilde{\mu}\vert)t\,)\exp(imvx)
\left\{ \exp(-i(\vert\tilde{\mu}\vert+\hbar {\bf k}^{2}/2m + k_{x}v)t\,)
 \exp(i{\bf k}{\bf x})\,u_{\bf k}\,\right\}
$$
It is easy to see that such a  collective  excitation,\, 
$\omega_{\rm ex}=\vert\tilde{\mu}\vert + \hbar {\bf k}^{2}/2m + k_{x}v$,\, 
can be interpreted 
as an exciton  with the energy $E_{g}+ \hbar^{2}\tilde{\bf 
k}^{2}/2m$, where $\hbar\tilde{k}_{j}=\hbar k_{j} + mv\delta_{1j}$.   
Then we can compare the condensate energy $E_{\rm o}(N_{\rm o})$  and 
the energy of the condensate with one outside excitation, 
$$
E_{\rm o}(N_{\rm o}-1) +  E_{\rm exc}(\tilde{k})+ E_{\rm ph}(k')\approx
E_{\rm o}(N_{\rm o})-\partial_{N}E_{\rm o}(N_{\rm o})+ (E_{g} +\hbar^{2}\tilde{k}^{2}/2m) 
+ \hbar c_{l}\vert k'\vert
\approx
$$
$$
\approx
E_{\rm o}(N_{\rm o})+(\hbar^{2}\tilde{k}^{2}/2m\,-\,mv^{2}/2) +  
3(\vert\tilde{\mu}\vert + (\nu_{0}/3)\Phi^{2}_{\rm o}\,) - 
 \frac{c_{l}^{2}+v^{2}}{c_{l}^{2}-v^{2}}\,\frac{\sigma_{0}^{2}}{\rho(c_{l}^{2}-v^{2})}\Phi_{\rm o}^{2}
+\hbar c_{l}\vert k'\vert    
$$
$$
 > E_{\rm o}(N_{\rm o})\,\, {\rm in}\,\, k\rightarrow 0 \,\,\,(k'\ne 0)\,\,\, {\rm limit}. 
$$ 
Note that the energy (and the momentum) of the  phonon part of the condensate changes after exciton emission. 
We assume that the transformation
$N_{\rm o}\rightarrow N_{\rm o}-1$ (or emission of an outside exciton)
 corresponds to the situation when  
the outside exciton and  the outside phonon(s) appear together, and
the phonon is emitted with the energy compensating 
the changement of $-\delta E_{\rm ph}$ in the phonon part of the condensate.   

However, the condensate collective excitations  
are uncoupled 
only in the $k\rightarrow 0$  limit, i.e. $\lambda= 2\pi/k \gg L_{0}$. 
It follows from the structure 
of Eqs. (\ref{eigfunction1})-(\ref{eigfunction3}) that the coupling 
between the  {\it outside} phonon and exciton branches 
is originated from the condensate 
``surface'' area, i.e. from the scale $L_{o}<\vert x \vert <  3L_{0}$. 
Indeed,  $\phi_{\rm o}(x)$ and $\phi_{\rm o}'(x)\simeq \pm \phi_{\rm o}(x)L_{0}^{-1}$ cannot be put
equal zero in this area, and  
the ``particle''-  and the ``hole''-type components 
of the exciton operators, namely
${\rm u}_{s}\sim {\rm e}^{ikx}$ and ${\rm v}^{*}_{s}\sim {\rm e}^{-ikx}$,
should be both different from zero and spatially  
modulated in the surface area, at least for the excitations with $\lambda  < 
2L_{0}$. 
We left this question for future investigations.

\subsection{Inside-Excitations}

To simplify the calculation of {\it inside}-excitation spectrum  
(see Eqs. (\ref{eigfunction1})-(\ref{eigfunction3})\,) 
we will use the semiclassical approximation \cite{Pit}.
In this approximation, the excitations can be labeled 
by the wave vector ${\bf k}$ in the co-moving frame, and the following
representation holds:
\begin{equation}  
{\rm u}_{s}({\bf x})= {\rm u}_{\bf k}({\bf x}){\rm e}^{ i\varphi_{\bf k}({\bf x})},\,\,\, 
{\rm v}_{s}({\bf x})= {\rm v}_{\bf k}({\bf x}){\rm e}^{i\varphi_{\bf k}({\bf x})},\,\,\,
 C_{s,j}({\bf x})= C_{{\bf k},j}({\bf x}){\rm e}^{ i\varphi_{\bf k}({\bf x})},
\label{semiclassical}
\end{equation}
where the phase $\varphi_{\bf k}({\bf x})\approx \varphi_{\rm o}+ {\bf kx}$, and ${\rm u}_{\bf k}({\bf x})$,
${\rm v}_{\bf k}({\bf x})$, and $C_{j,{\bf k}}({\bf x})$ are assumed to be  smooth functions of ${\bf x}$ in the inside condensate
area.
 Notice that the ${\bf k}$- and ${\bf x}$-representations are mixed here. This means 
that  
the operator nature
of the fluctuating fields is factually dismissed within the semiclassical approximation, 
However, the 
orthogonality 
relations between ${\rm u}_{s}$ and ${\rm v}_{s'}$, and, hence, 
between ${\rm u}_{\bf k}$ and ${\rm v}_{\bf k'}$   
come from the Bose commutation relations between 
the operators $\alpha_{s}$ and $\alpha^{\dag}_{s'}$ \cite{Fetter},\cite{Pit}.  
For example, Eq. (\ref{uvtransform1}) is modified as follows:    
\begin{equation}
 \delta{\psi}({\bf x},t)\simeq \!\int\!\frac{d{\bf k}}{(2\pi)^{3}} \,
{\rm u}_{\bf k}({\bf x}){\rm e}^{i\varphi_{\bf k}({\bf x})}
                            {\rm e}^{-i\omega_{\bf k}({\bf x})t}
\,+\,{\rm v}_{\bf k}^{*}({\bf x}){\rm e}^{-i\varphi_{\bf k}({\bf x})}{\rm e}^{ i\omega_{\bf k}({\bf x})t},
\label{uvtransform111}
\end{equation}
and the {\it inside}-excitation part  of the elementary excitation term in 
(\ref{ham3}),
\newline
$\sum_{s}...\approx \sum_{s,\,{\rm out}}+ \sum_{s,\,{\rm surf}} +\sum_{s,\,{\rm in}}...$, 
can be written in the form 
\begin{equation}
\sum_{s,\,{\rm in}}\hbar\omega_{s}\hat{\alpha}_{s}^{\dag}\hat{\alpha}_{s}\simeq
\int\!\frac{d{\bf k}\,d{\bf x}}{(2\pi)^{3}}\,\hbar\omega_{\bf k}({\bf x})\,n_{\bf k}({\bf x}). 
\label{semienergy}
\end{equation}
Note that the semiclassical energy $\hbar\omega_{\bf k}(x)$
of the  inside-excitation 
mode 
is supposed to be a smooth function of x, (i.e. at least as smooth as 
$\phi_{\rm o}^{2}(x)\approx {\rm const}$ in the ``inside'' approximation).  

Although the low-lying excitations cannot be properly described within the semiclassical approximation, 
we apply it here to calculate the low energy asymptotics 
of the spectrum. In fact, 
all the important properties of these excitations can be understood within this approach.

Substituting (\ref{semiclassical}) 
into Eqs. (\ref{eigfunction1})-(\ref{eigfunction3}), we transform these differential equations into the algebraic ones  
$(\,L(-{\bf k}^{2})=\hat{L}(\Delta \rightarrow -{\bf k}^{2})\,)$:
\begin{equation}
(L(-{\bf k}^{2}) - \hbar\omega_{\bf k})\,{\rm u}_{\bf k}({\bf x}) + \left(\nu_{0}\phi_{\rm o}^{2}({\bf x}) + 
2\nu_{1}\phi_{\rm o}^{4}({\bf x})\right)
{\rm v}_{\bf k}({\bf x}) +\sigma_{0}\phi_{\rm o}({\bf x})i{\bf k}{\bf C}_{\bf k}({\bf x})=0, 
\label{eigfunction01}
\end{equation}
\begin{equation}
\left(\nu_{0}\phi_{\rm o}^{2}({\bf x}) + 
2\nu_{1}\phi_{\rm o}^{4}({\bf x})\right){\rm u}_{\bf k}({\bf x})+
( L(-{\bf k}^{2}) +\hbar\omega_{\bf k})\,{\rm v}_{\bf k}({\bf x}) +
 \sigma_{0}\phi_{\rm o}({\bf x})i{\bf k}{\bf C}_{\bf k}({\bf x})=0
\label{eigfunction02}
\end{equation}
\begin{equation}
\rho^{-1}\sigma_{0}
\phi_{\rm o}(x)i k_{j}\,{\rm u}_{\bf k}({\bf x})
+\rho^{-1}\sigma_{0}
\phi_{\rm o}(x)ik_{j}\,{\rm v}_{\bf k}({\bf x})+
\left[(\omega_{\bf k}+vk_{x})^{2} - c_{l}^{2}{\bf k}^{2} \right]\,
C_{{\bf k},j}({\bf x})=0.
\label{eig} 
\end{equation}  
After some straightforward algebra, we can write out the equation for 
the exciton-phonon excitation spectrum:   
$$  
\left(\,(\omega_{\bf k}+vk_{x})^{2} - c_{l}^{2}{\bf k}^{2}\right)\times
$$
$$
\times \left[\, (\hbar\omega_{k})^{2}-
\left( L(-{\bf k})-
 \Bigl(\nu_{0}\phi_{\rm o}^{2}({\bf x}) + 
2\nu_{1}\phi_{\rm o}^{4}({\bf x})\Bigr)\right)
\left( L(-{\bf k})+
 \Bigl(\nu_{0}\phi_{\rm o}^{2}({\bf x}) + 
2\nu_{1}\phi_{\rm o}^{4}({\bf x})\Bigr)\right)
\,\right]=
$$
\begin{equation}
=\left(\,
 L(-{\bf k})-
 \Bigl(\nu_{0}\phi_{\rm o}^{2}({\bf x}) + 
2\nu_{1}\phi_{\rm o}^{4}({\bf x})\Bigr)
\,\right)
\,\frac{2\sigma_{0}^{2}}{\rho c_{l}^{2}}\phi_{\rm o}^{2}(x)\,(c_{l}^{2}{\bf k}^{2}).
\label{spectrum0}
\end{equation}

Note that within the semiclassical description of the {\it inside}-excitations, the low energy
limit means $k\rightarrow k_{0}$ where $k_{0}$ is the momentum cut-off,
$$
 (\hbar^{2}/2m){\bf k}_{0}^{2}\simeq \vert\tilde{\mu 
}\vert=(\hbar^{2}/2m)L_{0}^{-2}. $$
The inequality $k>k_{0}\simeq L_{0}^{-1}$ ensures the function 
$\hbar\omega_{\bf k}({\bf x})$
in (\ref{spectrum0}) to be real and positive.
Indeed, only the excitations with the wave lengths $\lambda < 
(2\sim 3 )L_{0}$ can be considered 
as the {\it inside} ones. 
The presence of the ``hard core'' terms,  
${\rm const}\,\nu_{1}\phi_{\rm o}^{4}(x)$, 
leads to a slight renormalization of the value of the momentum cut-off.   
However, this renormalization does not factually change  the characteristic 
properties of the possible solutions of Eq. (\ref{spectrum0}). 
We will mark the ``hard core'' terms 
by $\epsilon_{+}>0$.     
For example, in the low energy limit, 
$$ 
k \approx k_{0} + \delta k,\,\,\,\,    
\delta k \rightarrow 0,
$$ 
the exciton part of the l.h.s.
of Eq. (\ref{spectrum0})\,-- i.e. the formula inside the square brackets -- can be 
reduced to the form: 
\begin{equation} 
(\hbar\omega_{\bf k})^{2} - \left(
\frac{\hbar^{2}({\bf k}^{2}-{\bf k}_{0}^{2})}{2m} +
F(x)+ \epsilon_{+}\right)\, 
\left( \frac{\hbar^{2}({\bf k}^{2}-{\bf k}_{0}^{2})}{2m} +
F(x)+2\nu_{0}\phi^{2}_{\rm o}(x) + \epsilon_{+}\right), 
\label{excitonpart}
\end{equation}
where $F(x)=\left(\sigma_{0}^{2}/\rho (c^{2} - v^{2})\,-\, \nu_{0}\right)
(\Phi_{\rm o}^{2} - \phi_{\rm o}^{2}(x)\,)$, and 
the following two estimates hold:
$F(x)\simeq 2\vert\tilde{\mu}\vert (2x/L_{0})^{2}$ at 
$x\sim 0$ and $F(x)\simeq 2\vert\tilde{\mu}\vert$ at $x  
>\pm 
2L_{0}$. 

There are two different types of the inside-excitations, the longitudinal
excitations and the transverse ones. The 
later have the wave vectors ${\bf k}$ perpendicular to the 
$x\,(v)$-direction. 
Although  Eq. (\ref{spectrum0})
 can be solved exactly for the transverse excitation spectrum \cite{book},
taking into account the coupling term  
changes the values of excitation energies slightly, and  
the excitations 
can be approximately considered as of the pure exciton or phonon types.
Then we have the acoustic phonon dispersion low for
the phonon branch and the following
spectrum for the exciton branch:  
$$
(\hbar\omega_{{\rm ex},\,k_{\perp} })^{2}\simeq 
\left( \frac{\hbar^{2}}{2m}(k^{2} 
- k_{0}^{2}) +\epsilon_{+} \right)\, 
\left( \frac{\hbar^{2}}{2m}(k^{2} 
- k_{0}^{2}) + 2\nu\phi_{\rm o}^{2}(x) 
+\epsilon_{+}\right)\approx 
$$
\begin{equation}
\approx \frac{\hbar^{2}}{2m}(k^{2} 
- k_{0}^{2})\,2\nu\phi_{\rm o}^{2}(x) 
+2\nu\phi_{\rm o}^{2}(x)\epsilon_{+} 
< \left( \,\vert\tilde{\mu}\vert +(\hbar^{2}/2m )k_{\perp}^{2}\,\right)^{2}. 
\label{exciton11}
\end{equation}
The smooth function $\omega_{k_{\perp}}(x)>0$ has a gap when
$k \rightarrow k_{0}$, but 
 unlike the case of the outside excitations, the gap value is determined by the ``hard 
core'' repulsion term and is much less than $\vert \mu \vert$. 
Furthermore, if we let (formally) the x coordinate in  (\ref{exciton11}) change in the area of
$\vert x \vert > L_{0} $, the dispersion low $\hbar\omega_{{\rm ex},\, k_{\perp}}(x)$ reproduces
the outside excitation asymptotics, $\vert\tilde{\mu}\vert +(\hbar^{2}/2m )k_{\perp}^{2} $. 
However, inside the condensate,
we obtain a strong deviation of the collective excitation spectrum from the simple excitonic 
one.

In the case of the longitudinal excitations, the mode interaction is 
non-negligible in the low energy limit. 
For instance, 
the ``exciton'' root of Eq. (\ref{spectrum0}) with the nonzeroth r.h.s. 
exists if 
$k_{x}>2.5\,L_{0}^{-1}$.
Therefore there are no distinct exciton and phonon
modes, and the cases $k_{x}>0$ and $k_{x}<0$ are different because of
different position of the ``bare'' phonon root on the energy axis. 
Then, on the energy axis $\hbar \omega$, the modified phonon spectrum is located higher 
than the phonon frequencies and the modified exciton spectrum
is located lower then $  \hbar\omega_{{\rm ex},\,k_{x}}^{({\rm o})}$.
Yet, like the case of transverse excitations, the same inequality and the same 
(formal) asymptotics are 
valid for the lower branch of the spectrum:    
$$  
0<\omega_{k_{x}}< \vert\tilde{\mu}\vert+ (\hbar^{2}/2m)k_{x}^{2}.  
$$    
The approximate formulas for the longitudinal spectrum are too cumbersome 
to be presented here. However,
the phonon component of the excitonic longitudinal excitation, $C_{k,\,1}(x) \ne 0$, can be found  
approximately from Eq. (\ref{eig}) by use of the Bogoliubov form for the wave functions ${\rm 
u}_{k}^{2}(x)$   and ${\rm v}_{k}^{2}(x)$ \cite{Pit}:
\begin{equation}
{\rm u}^{2}_{k}(x) \approx  \left(\frac{1}{V_{\rm eff}}\right)\frac{ L(-k^{2}) 
+\hbar\omega_{k} }{2\hbar\omega_{k}},
\,\,\,\,{\rm v}^{2}_{k}(x) \approx \left(\frac{1}{V_{\rm eff}}\right)
\frac{ L(-k^{2}) -\hbar\omega_{k}}{2\hbar\omega_{k}},
\label{mmm}
\end{equation}
where $L(-k^{2})\approx  \hbar^{2}(k^{2} 
- k_{0}^{2})/2m \,+ \nu\phi_{\rm o}^{2}(x) + \epsilon_{+}$. The effective condensate 
volume $V_{\rm eff} \simeq  4SL_{0} $ is used to normalize 
the u- and v-wave functions of the inside 
excitations, $\int d{\bf r}(\vert {\rm u}_{k}\vert ^{2} - \vert {\rm v}_{k}\vert ^{2})=1$. 
Subsiquently, we get 
\begin{equation}
C_{k,\, 1}(x)\approx \frac{\rho^{-1}\sigma_{0}\phi_{\rm o}(x)i k_{x}\,\bigl({\rm u}_{k}(x) +{\rm v}_{k}(x)\,\bigr)}
{ c_{l}^{2}{ k}^{2} - (\omega_{k_{x}}+vk_{x})^{2} }.
\label{mmm1}
\end{equation}
One can use the zeroth approximation, Eq. (\ref{exciton11}), for  $\omega_{ k_{x} }$ in (\ref{mmm}),(\ref{mmm1}).  
Then we can roughly estimate the maximum value of $\vert C_{k}(x) \vert^{2}$ in the low energy limit 
($k \rightarrow 2.5 L_{0}^{-1}$, $\omega \ll c \vert k_{x} \vert $): 
\begin{equation}
\vert C_{k}\vert^{2} \simeq \left( \vert\tilde{\mu}\vert /\rho(c_{l}^{2} - v^{2})V_{\rm eff}\,\right)L_{0}^{2} <\!\!<\!\!<
L_{0}^{2}.
\end{equation}


To investigate the stability of the moving condensate in 
relation to the creation of inside 
excitations, we can calculate the energy 
of the condensate with the {\it one} inside excitation described by 
the  following set: $k$, $\omega_{k}$, ${\rm u}_{k}$ and ${\rm v}_{k}$, and $C_{k}$. 
\,Although such an excitation
was defined  in the co-moving frame, calculations should be done in the laboratory frame. 
Returning to the lab frame, we represent
the exciton and phonon field functions as follows:   
 \begin{equation}
\phi_{\rm o}(x-vt,\,t) \rightarrow \phi_{\rm o}(x-vt,\,t) + \exp\bigl(-i(E_{g} + mv^{2}/2 
-\vert \tilde{ \mu }\vert) t\,\bigr)\exp(imvx)\,\delta\tilde{\Psi}({\bf x}, t),
\label{urfin}
\end{equation}
$$
\delta\tilde{\Psi}({\bf x}, t)=
{\rm u}_{k}(x-vt)e^{ i{\bf k}{\bf x}}e^{-i(\omega_{k}+k_{x}v)t } +
{\rm v}_{k}(x-vt)e^{- i {\bf k}{\bf x}}e^{+i(\omega_{k}+k_{x}v)t},  
$$
\begin{equation}
u_{\rm o}(x-vt) \rightarrow u_{\rm o}(x-vt) + 
C_{k}(x-vt)\exp(i{\bf k}{\bf x})\exp\bigl(-i(\omega_{k}+k_{x}v)t\,\bigr)+ {\rm c.c.}\,.
\label{urfin1}
\end{equation}
As the inside excitation is considered as an fluctuation,
the number of particles in the condensate  and its energy can be estimated as 
$N_{\rm o} - \int d{\bf x}\, \delta\psi^{\dag}\delta\psi$ and 
$E_{\rm o}(N_{\rm o}) - \partial_{N}E_{\rm o}\,\int d{\bf x}\,\delta\psi^{\dag}\delta\psi$,
respectively.

The zeroth component 
of the energy-momentum tensor can be represented in the form   
$$
{\cal T}_{0}^{0}={\cal T}_{0}^{0}(\phi_{\rm o}, \,u_{\rm o})+ 
{\cal T}_{0}^{0\,(2)}(\delta\Psi^{\dag}, \delta\Psi, \delta u\,\vert\,\phi_{\rm o}, \,u_{\rm o}),
$$
where the first part gives the condensate energy $E_{\rm o}$ and the second
part will give the energy of the inside excitation, $E_{\rm in}$. 
After substitution of (\ref{urfin}),(\ref{urfin1}) 
into $E_{\rm in}=\int d{\bf x}\,{\cal{T}}_{0}^{0\,(2)}$,  
 it can be rewritten as follows: 
\begin{equation}
E_{\rm in}=\int d{\bf x}(E_{g}+ mv^{2}/2 - \vert\tilde{\mu}\vert)\delta\tilde{\psi}^{\dag}
\delta\tilde{\psi} + \int d{\bf x}\,\hbar(\omega_{k}+k_{x}v)(\vert {\rm u}_{k} \vert^{2} -
\vert {\rm v}_{k}\vert^{2}) + 2\rho\vert C_{k} \vert ^{2}(\omega_{k} + vk_{x})^{2}.
\label{bbb}
\end{equation}
The first term appearing in (\ref{bbb}),
 $\delta\tilde{\psi}^{\dag}\delta\tilde{\psi}\rightarrow 
\vert {\rm u}_{k}\vert^{2}+\vert {\rm v}_{k} \vert^{2}$, vanishes in the following formula
for the total energy: 
$$
E_{\rm o}+E_{\rm in}\approx E_{\rm o}(N_{\rm o}) + (2\vert\tilde{\mu}\vert + \nu_{0}\Phi_{\rm o}^{2}) 
\int d{\bf x}\,\delta\tilde{\psi}^{\dag}
\delta\tilde{\psi} + \int d{\bf x}\,\hbar(\omega_{k}+k_{x}v)(\vert {\rm u}_{k} \vert^{2} -
\vert {\rm v}_{k}\vert^{2})\,-
$$  
\begin{equation}
-\frac{c_{l}^{2}+v^{2}}{c_{l}^{2}+v^{2}} \frac{\sigma_{0}^{2}}{\rho(c_{l}^{2}-v^{2})}
\Phi^{2}_{\rm o} \int d{\bf x}\,\delta\tilde{\psi}^{\dag}
\delta\tilde{\psi} + \int d{\bf x}\,2\rho\vert C_{k} \vert ^{2}(\omega_{k} + vk_{x})^{2}.
\label{stable}
\end{equation}
Here the last two terms compensate each other approximately, and  
all the interesting effects come form the exciton part of (\ref{stable}).
Although $\omega_{k}+k_{x}v$ can be negative, its negative value can be compensated
by the  term $2\vert\tilde{\mu}\vert + \nu_{0}\Phi_{\rm o}^{2}$ 
if the velocity of the condensate is close to $c_{l}$ or the exciton concentration is high enough.
Therefore, 
there is a second critical velocity $v_{c}$ in the theory. 
If the velocity of the condensate is less then the velocity $v_{c}$ and $v_{\rm o}< v_{c} < c_{l}$, 
the condensate is unstable in  
relation to creation of the inside excitations, i.e. $E_{\rm o} +E_{\rm in}(k_{x},v)< E_{\rm o}$ in the lab frame.  
To estimate the value of $v_{c}$, we solve the following equation:
$$
(3 \sim 5)\hbar v L_{0}^{-1}(v) \simeq \left( \sigma_{0}^{2}/(c_{l}^{2} - v^{2})\rho \right)\Phi_{\rm o}^{2}(v), 
$$
which can be reduced to $(3 \sim 5)\hbar v \simeq ( \sigma_{0}^{2}/(c_{l}^{2} - v^{2})\rho)\,(N_{\rm o}/2S)$. 
For example, for the packets with the exciton concentration of $n\simeq 10^{14}\sim 10^{15}$\,\,cm$^{-3}$, the estimate
is as follows:
\begin{equation}
\frac{ c_{l}-v_{c} }{ c_{l} } \simeq \frac{ \sigma_{0}^{2} }{ \rho c_{l}^{2} }\,\frac{ N_{\rm o} }{2S}\,
\frac{0.1}{\hbar c_{l} }\simeq 0.1 \sim 0.3.
\end{equation}

\section{Interference between Two Moving Packets}

There are at least two interesting problems that can be examined  in the
framework of the proposed model (Eqs. (\ref{eq11},\ref{eq22})\,). 
The first problem is the
investigation of the condensate steady-state and its stability, calculation of the 
low-lying excitation spectrum, etc.. (Some part of this program was presented in the previous sections.) 
The second  one is the investigation of interference between two
moving packets \cite{interfer}. In this case the problem is
essentially nonstationary. 
Indeed, the amplitude and the shape of the resultant  moving packet are 
expected to change in time \cite{excitoner}.        
These effects, however, can be considered theoretically  
by a numerical solving  
of Eqs. (\ref{eq11}),(\ref{eq22}) with the proper initial conditions.  
(We will disregard the influence of noncondensed particles, the condensate depletion  and  
nonequilibrium phonons on the dynamic processes being considered.)
For example,
if two ``input'' packets have the same velocity 
(${\bf v}_{1} = {\bf v}_{2}={\bf v}$) and shape,  
we can  write the initial conditions in the  form       
 \begin{equation}
\psi_{0}({\bf x}, t=0)\cdot {\bf u}_{0}({\bf x},t=0)=
\phi_{\rm o}({\bf x})\cdot {\bf q}_{\rm o}({\bf x}) +  {\rm 
e}^{i\delta\varphi}
\phi_{\rm o}({\bf x} + {\bf v}\tau)\cdot {\bf q}_{\rm o}({\bf x}+{\bf 
v}\tau), 
\label{conditions}
\end{equation}
where $\delta\varphi = {\rm const}$ is the 
macroscopical phase  difference between the two ``input'' condensates,  and 
$\tau = {\rm const}$  is the time delay between them.

In the simplest (quasi-1D) model, the following equations govern the dynamics
of the two ``input'' packets: 
\begin{equation}
(\,i\hbar\partial_{t} + mv^{2}/2\,)\psi_{0}(x,t)= \bigl(-\frac{\hbar^{2}}{2m }\partial^{2}_{x}+
\nu_{0}\vert\psi_{0}\vert^{2} + \nu_{1}\vert\psi_{0}\vert ^{4}\bigr)\,\psi_{0}(x,t) + 
\sigma_{0}\,\partial_{x} u_{0}(x,t)\,\psi_{0}(x,t)
\label{timedep1D}
\end{equation}
\begin{equation}
\bigl(\partial_{t}^{2}- (c_{l}^{2} - v^{2})\partial_{x}^{2}- 2v\partial_{t}\partial_{x} 
 \,) u_{0}( x,t)=
\rho^{-1}\sigma_{0}\partial_{x}\vert\psi_{0}\vert^{2}( x,t).
\label{timedep1DD}
\end{equation}
Then the initial conditions (\ref{conditions}) can be written in the explicit 1D form by using  the
exact solution (\ref{Soliton}) of the model (\ref{1Deq},\,\ref{11Deq}). Note that the amplitudes 
of $\phi_{\rm o}(x)$  and $\partial u_{\rm o}(x)$
are defined from the normalization condition  and depend on
the values of $v$ and $N_{\rm o}$, and, hence,
the amplitudes of the ``input'' condensates in (\ref{conditions})
have the same values.

Some predictions of the form of the expected solution can be made easily.
Indeed, the shape and other characteristics of the steady-state
solution of (\ref{timedep1D}),\,(\ref{conditions}) will depend mainly on the value of 
the parameter $x_{0}/L_{0}$ where $ x_{0}=v\tau$ and $L_{0}$  is 
the characteristic condensate width.      
If $x_{0}/L_{0} < 1\sim2$, the nonlinear interaction between the packets
plays an important role in the process of total wave function  formation.
(Notice that the lower limit of $x_{0}$, \,$x_{0}^{*}=\tau^{*}v$, is defined by the time
scale of formation of a condensate wave function.) 
It is reasonable to assume that in the limit of strong  interaction between 
condensates, the $N_{\rm o}+N_{\rm o}$ excitons can form the {\it single} condensate
wave function (\ref{Soliton}) in the steady-state regime. Then this 
wave function can be written (in the laboratory frame) as follows:  
$$
\psi_{0}(x,t)\cdot u_{0}(x,t)\simeq
\exp\bigl(- i(\tilde{\mu}(2N_{\rm o},\tilde{v})+ m\tilde{v}^{2}/2)t\,\bigr)
\exp(im\tilde{v}x)\times
$$
$$
\times\exp(i\tilde{\varphi})\phi_{\rm 
o}(x-\tilde{v}t;\,2N_{\rm o})\cdot q_{\rm o}(x-\tilde{v}t;\,2N_{\rm o}). $$   
As the dynamic equations (\ref{timedep1D}),(\ref{timedep1DD}) conserve the energy,    
$$
E_{\rm in}(N,v;\,x_{0}/L_{0})=E_{\rm out}(2N,\tilde{v}),
$$
$\tilde{v}$ cannot be equal to $v$ in theory. Moreover, if the 
parameter $x_{0}/L_{0}$ is small enough, it can be only the approximate 
equality,  $\tilde{v}\approx v$. 

In the case of  $x_{0}/L_{0} \gg 1$, one can disregard the influence of the 
mutual nonlinear interaction on the 
dynamics of the packets. In this approximation, the packet moving in the crystal can be
modeled by the following formula:  
$$ \psi_{0}(x,t)\cdot u_{0}(x,t)\simeq
\exp\bigl(-i(\tilde{\mu}(N_{\rm o})-mv^{2}/2)t\,\bigr)
\phi_{\rm o}(x;\,N_{\rm o})\cdot q_{\rm o}(x;\,N_{\rm o})\,+\,
$$
$$
+\exp(i\delta\varphi)\exp\bigl(-i(\tilde{\mu}(N_{\rm o})-mv^{2}/2)t\,\bigr)
\phi_{\rm o}(x+x_{0};\,N_{\rm o})\cdot q_{\rm o}(x+x_{0}; N_{\rm o}).
$$

An interesting and noninvestigated case in the interference problem  is
the condensate dynamics after posing nonsymmetric initial 
conditions. In fact, the amplitude and the velocity of the ``input'' 
packets can be different, for example, $N_{2} > N_{1}$ and $v_{2}>v_{1}$,
${\bf v}_{2}\|{\bf v}_{1}$.
We use here the experimental result \cite{Fortin}
that at $v>v_{0}$, the 
velocity of the condensate depends on the laser power or,
equivalently, on the initial concentration of excitons, i.e. $v=v(N)$.
(Note that in theory the exciton amplitude $\Phi_{\rm o}$ is the function of 
$N_{\rm o}$  and  $v$.)

If the exciton concentration in the first packet, $n_{1}$, is close to the 
Bose condensation threshold 
value and the exciton concentration in
the second packet, $n_{2}$, can be made $\gg n_{1}$,
the velocity difference between condensates can 
reach  
$(0.2\sim 0.3)c_{l}$.  
Then, in the reference frame moving with the first packet, the initial
conditions can be taken as the following: 
$$
\psi_{0}({\bf x},t=0)\cdot {\bf u}_{0}({\bf x},t=0)= 
\phi_{\rm o}({\bf x};N_{1})\cdot {\bf q}_{\rm o}({\bf x};N_{1})\,+\,
$$
\begin{equation}
+\exp(i\delta\varphi )\exp(im\,\delta{\bf v}\,{\bf x})\,
\phi_{\rm o}({\bf x}+{\bf x}_{0};N_{2})\cdot {\bf q}_{\rm o}
({\bf x}+{\bf x}_{0};N_{2}),
\label{nonsym}
\end{equation}
where $\delta {\bf v} = {\bf v}_{2}-{\bf v}_{1}$, $x_{o}=v_{1}\tau $,
and the second packet moves in this frame of reference.
In the case of such the initial conditions, the 
regime of 
strong nonlinear interaction between the condensates is unavoidable. 
Following the logic of the soliton theory, 
we speculate that the steady-state solution may consist of two packets 
moving with different velocities and with different exciton concentrations.  
Roughly speaking, the ``input'' packets could 
exchange their places, i.e. the {\it both} packets 
survive 
after collision, and the first packet arrives at the
``detecting'' boundary of a crystal after the second one: 
$$ 
\psi_{0}(x,t)\cdot u_{0}(x,t)\simeq 
\exp\bigl(-i(\tilde{\mu}(N_{1})- mv_{1}^{2}/2)t\,\bigr)\,\phi_{\rm 
o}(x;N_{1})\cdot q_{\rm o}(x;N_{1})\,+
$$
$$
+ \exp\bigl(-i(\tilde{\mu}(N_{2})-mv_{1}^{2}/2\,+m\delta v^{2}/2)t\,\bigr)
\exp(i\delta\tilde{\varphi})\exp(im\,\delta v\,x)\times 
$$
$$
\times \phi_{\rm o}
(x+x_{0}-\delta v\,t; N_{2})\cdot q_{\rm o}(x+x_{0}-\delta v\,t; N_{2}).
$$
However, the hypothesis about the solitonic character of packet 
collisions in 3D needs both numerical and experimental evidence. 

\section{Conclusion}
 
In this study, we considered a model within which  
the inhomogeneous excitonic condensate with a nonzero momentum can be investigated. 
The important physics we include in our model is 
the exciton-phonon interaction
and the appearance of 
a coherent part of the crystal displacement field that makes the moving condensate
of the exciton-phonon one. Then, the condensate wave function and its energy
can be calculated exactly in the simplest quasi-1D model.
We believe that the transport and other unusual properties 
of the coherent para-exciton
packets in Cu$_{2}$O
can be described in the framework of the proposed model properly generalized 
to meet more realistic conditions. 

As the exciton-phonon interaction is very important 
in any processes involving excitons in lattices \cite{somebook},
we can speculate about  
the possibility of use of piezoelectrical transducers to pump
acoustic waves into the system condensate\,$+$\,lattice. Moreover, the transducers 
could be used to register the phonon part of the coherent packet 
in experiments in which  the condensate is formed by optically inactive 
excitons and phonons.

We showed that there are two critical velocities in the theory, $v_{0}$ and $v_{c}$. 
The first one, $v_{0}$, comes from the renormalization of two particle exciton-exciton interaction due to phonons,
and the inhomogeneous soliton state can be formed if $v>v_{0}$.
The second one, $v_{c}$, comes from use of Landau arguments \cite{Fetter} for investigation of the dynamical
stability\,/\,instability of the moving condensate. Within 
the semiclassical approximation for the condensate excitations, we found the condensate is unstable if $v<v_{c}$.       
It is interesting to discuss the possibility of observation of such an instability when the condensate can be formed
in the inhomogeneous state with $v \ne 0 $, but with $ v_{0} <v< v_{c}(n,v)$. 
Such a coherent packet has to disappear during
its move through a single pure crystal used for experiments. As the shape of the moving packet depends on time,
the form of the registered signal may depend on the crystal length
changing from the solitonic to the standard diffusion density profile.

We did not concentrate on detailed investigation of excited states of the moving
exciton-phonon condensate in this study.  
First, the possibility of their observation
is an unclear question itself.
Second, the stability 
conditions of 
the  moving condensate --  in relation to the creation of condensate 
excitations -- 
can be derived from the low energy asymptotics of the excitation spectra at $T\ll T_{c}$.
However, the stability problem is not without difficulties
\cite{book},\cite{Rou}.  
One can easily imagine the situation when
the condensate moves in a very high quality crystal, but with some impurity region carefully
prepared in the middle of the sample. Then the impurities
could bound the noncondensed excitons, which always accompany the condensate, and could mediate, for instance, the
emission of the outside excitations. The last process may lead to depletion of the condensate and, perhaps,
some other observable effects, such as damping, bound exciton PL, etc.. 
On the other hand, the 
inside 
excitations could manifest themselves
at $T\ne 0$ by the effective 
enlargement of the packet length, $L_{0}\rightarrow L_{\rm eff}$, $T\ne 0$, or by interaction
with  external acoustic waves.    

\section{Acknowledgements}

One of the authors (D.R.) thanks I.~Loutsenko for useful discussions 
and E.~Benson and E.~Fortin for  
providing the results  
of their work before publishing.


\begin{thebibliography}{100}

\bibitem{review}
J. P. Wolfe, J. L. Lin,
D. W. Snoke and by A. Mysyrowicz in {\it Bose-Einstein
Condensation}, edited by
A. Griffin, D. W. Snoke and S. Stringari (Cambridge University Press,
Cambridge, 1995).

\bibitem{Butov} 
L. V. Butov, A. Zrenner, G. Abstreiter, G.~Bohm, and
 G.~Weimann, Phys. Rev. Lett {\bf 73}, 304, (1994); Physics--Uspekhi {\bf 39},
751 (1996).

 
\bibitem{Lin}
J. L. Lin, J. P. Wolfe, Phys. Rev. Lett. {\bf 71}, 122
(1993).

\bibitem{Goto}
T. Goto, M. Y. Shen, S. Koyama, T. Yokouchi, Phys. Rev. B
{\bf 55}, 7609 (1997). 
 
\bibitem{Fortin}
E. Fortin, S. Fafard, A. Mysyrowicz, Phys. Rev. Lett.
{\bf 70}, 3951 (1993).

\bibitem{Benson}
E. Benson, E. Fortin, A. Mysyrowicz, Phys. Stat. Sol. B {\bf 191}, 345 (1995);
Sol. Stat. Comm. {\bf 101}, 313, (1997). 

\bibitem{Hanamura1} 
E. Hanamura, Sol. Stat. Comm. {\bf 91}, 889 (1994);
\newline
J.~Inoue, E.~Hanamura, Sol. Stat. Comm. {\bf 99}, 547 (1996);

\bibitem{Tichodeev}
A. E. Bulatov, S. G. Tichodeev, Phys. Rev. B {\bf 46}, 15058 (1992);
\newline
G. A. Kopelevich, S. G. Tikhodeev, and N. A. Gippius, JETP {\bf 82}, 1180 (1996).

\bibitem{Loutsenko}
I. Loutsenko, D. Roubtsov, Phys. Rev. Lett. {\bf 78},
3011 (1997).

\bibitem{boser}
A.~Imamo\=glu and R.~J.~Ram, Phys.  Lett.  A {\bf 214},
193 (1996);
\newline
W. Zhao, P. Stenius, A.~Imamo\=glu, Phys. Rev. B {\bf 56}, 5306 (1997). 

\bibitem{Kavoulakis}
G.~M.~Kavoulakis, G.~Baym, and J.~P.~Wolfe, Phys. Rev. B {\bf 
53}, 7227 (1996);
\newline
G.~M.~Kavoulakis, Y.-C. Chang, and G.~Baym, Phys. Rev. B {\bf 
55}, 7593 (1997);

\bibitem{discussion}
S. G. Tichodeev, Phys. Rev. Lett. {\bf 78}, 3225 (1997);
\newline A.~Mysyrowicz, {\it ibidem}, 3226 (1997).

\bibitem{Ivanov}
A. L. Ivanov, C. Ell, and H. Haug, Phys. Rev. B {\bf 57}, 9663 (1998).

\bibitem{Schmitt-Rink} S.~Schmitt-Rink, D.~S.~Chemla, and
D.~A.~B.~Miller, Adv. Phys. {\bf 38}, 89 (1989).

\bibitem{Keldysh}
L.~V.~Keldysh and A.~N.~Kozlov, Sov.  Phys.  JETP
{\bf 27}, 521 (1968);
\newline
E. Hanamura, H. Haug, Phys. Rep. C {\bf 33}, 209 (1977).

\bibitem{Popov}
V. N. Popov, {\it Functional Integrals and Collective Modes}, (Cambridge University Press, 
N.Y., 1987).

\bibitem{miscellaneous} 
M. F. Miglei, S. A. Moskalenko, and A. V. Lelyakov, Phys. Stat. Sol. {\bf 35}, 389
(1969);
\newline V. M. ~Nandkumaran and K. P. ~Sinha, Zeit. f\"ur Phys. B {\bf 22}, 173 (1975).

\bibitem{Griffin}
A.~Griffin, Phys. Rev. B {\bf 53}, 9341 (1996).

\bibitem{Fetter} 
Al. L. Fetter and J. D. Waleska, {\it Quantum Theory of Many-Particle 
System}, (McGrav-Hill, New York, 1971).

\bibitem{Pit}
S. Giorgini, L. P. Pitaevskii, S. Stringari, Phys. Rev. A  {\bf 54}, 4633 (1996);
\newline
F. Dalfovo, S. Giorgini, L. P. Pitaevskii, and S. Stringari, submitted to
Rev. Mod. Phys.; cond-mat/9806038. 

\bibitem{Hopfield} 
J.J. Hopfield, Phys. Rev. {\bf 112}, 1555 (1958).

\bibitem{book}  
S. A. Moskalenko, D. W. Snoke,
{\it Bose Condensation of Excitons and Coherent Nonlinear Optics}, (Cambridge University Press, in press).

\bibitem{interfer} 
E.~Benson, E.~Fortin, B. Prade, and A.~Mysyrowicz, to appear in Europhys. Lett..

\bibitem{excitoner}
A.~Mysyrowicz, E.~Benson, and E.~Fortin, 
Phys. Rev. Lett.  {\bf 77}, 896 (1996).

\bibitem{Rou}
D. Roubtsov, Y. L\'epine, to appear in Phys. Lett. A.; cond-mat/9807023.

\bibitem{somebook} 
H. Stolz, {\it Time-Resolved Light Scattering from Excitons}, 
(Springer-Verlag, 1994).

\end{thebibliography}
\end{document}